\documentclass[12pt,a4paper]{scrartcl}

\usepackage{float}
\usepackage{graphicx}
\usepackage{alg}
\usepackage{cite}
\usepackage{times}

\usepackage{amsfonts}
\usepackage{amssymb}
\newcommand{\disave}[1]{\left\langle #1 \right\rangle}
\newcommand{\bigo}{\mathcal{O}}
\newcommand{\np}{$\mathcal{NP}$}
\newcommand{\p}{$\mathcal{P}$}
\newcommand{\problem}[1]{{\sc #1}}
\newcommand{\probdef}[2]{
  \begin{quote}
    {\sc #1:} #2
  \end{quote}
}

\title{Computational Complexity for Physicists\footnote{published in {\it Computing in Science \& Engineering} {\bf 4} 31--47 (May/June 2002)}}

\author{S.~Mertens\thanks{stephan.mertens@physik.uni-magdeburg.de}\\
  \small Inst.~f.~Theor.~Physik, Otto-von-Guericke Univ.,
  PF 4120, D-39016 Magdeburg, Germany}

\date{}

\begin{document}

\maketitle

\begin{abstract}
{\bf Abstract:} These lecture notes are an informal introduction 
to the theory of computational complexity and its links to quantum
computing and statistical mechanics.

\end{abstract}

\section{Introduction}

\subsection{Motivation}

\label{sec:motivation}

Compared to the traditionally close
relationship between physics and mathematics, an exchange of ideas and
methods between physics and {\em computer science} barely takes
places.  The few interactions that go beyond Fortran programming and
the quest for faster computers were often successful and provided
surprising insights in both fields. This holds particularly for the
mutual exchange between statistical mechanics and the theory of 
computational complexity.

The branch of theoretical computer science known as computational
complexity is concerned with classifying problems according to the
computational resources (CPU time, memory) required to solve them.
This has lead to precisely defined notions of {\em tractable} and {\em
  intractable} problems. It appears as if this notions can be
transferred to the problem of {\em analytical solubility} of models
from statistical physics, explaining to some extent why for example
the Ising model is exactly soluble in two dimensions, but not in three
(sec.~\ref{sec:ising}).

The success of the theory of computational complexity is based on a
pessimistic attitude: a problem's tractability is defined according to
the worst possible instance. Quite often this worst case scenario
differs considerably from the typical case, averaged over a reasonable
ensemble of instances. A common observation is that hard problems are
typically easy to solve. To get real hard instances, the parameters of
the underlying ensemble must be carefully tuned to certain critical
values. Varying the parameters across the critcal region leads to abrupt
changes in the typical complexity---very similar to the abrupt changes
associated with phase transitions in physical systems. Phase
transitions in physical {\em and} computational systems are best studied within the
framework of statistical mechanics (sec.~\ref{sec:phase-transition}).
  
Apart from phase transitions, statistical mechanics offer means for the general
probabilistic analysis of computational problems.
The usual way is to formulate an optimization problem as a spin glass
and analyze the low temperature properties of the latter. This `physical' approach
often yield results which go beyond the results obtained by traditional methods
(sec.~\ref{sec:prob-analysis}).

Another exciting link between physics and computer science is provided by
{\em quantum computing}. There is some theoretical evidence that computers
using quantum systems as computational devices are more powerfull than
computers based on classical devices. The hope is that problems which
are intractable on a classical computer become tractable when put on
a quantum computer. Results like Shor's celebrated quantum algorithm for factorization 
nurture this hope, but a real breakthrough is still missing. Obviously some knowledge of 
computational complexity helps to understand the promises and limitations of quantum computers
(sec.~\ref{sec:quantum}).

These notes are directed at physicists with no or little knowledge
of computational complexity. Of course this is not the first
informal introduction into the field, see e.g.\ \cite{lewis:papadimitriou:78,hartmann:99}
or the corresponding chapters in textbooks on 
algorithms \cite{clr:algorithms,papadimitriou:steiglitz:82}.
For a deeper understanding of the field you are referred to the classical textbooks
of Garey and Johnson \cite{garey:johnson:79} and Papadimitriou \cite{papadimitriou:94}.

\subsection{The measure of complexity}

\label{sec:measure}

\subsubsection{Algorithms}

The computational complexity of a problem is a measure of the
computational resources, typically time, required to solve the
problem. What can be measured (or computed) is the time that a
particular {\em algorithm} uses to solve the problem.  This time in
turn depends on the implementation of the algorithm as well as on the
computer the program is running on. The theory of computational
complexity provides us with a notion of complexity that is largely
independent of implementational details and the computer at hand. Its
precise definition requires a considerable formalism, however. This is
not surprising, since it is related to a highly non trivial question
that touches the fundaments of mathematics: {\em What do we mean by
  saying a problem is solvable?} Thinking about this question leads to
G\"odel's incompleteness theorem, Turing machines and the Church
Turing Thesis on computable functions.  See \cite{penrose:89} for an
entertaining introduction into these topics written by a physicist.

Here we will adopt a more informal, pragmatic point of view. A problem
is solvable if it can be solved by a computer program written in your
favourite programming language.  The running time or {\em time
  complexity} of your program must then be defined with some care to
serve as a meaningful measure for the complexity of the problem.

\subsubsection{Time complexity}

In general the running time depends on the size of the problem and on
the specific input data, the instance.  Sorting $1000$ numbers takes
longer than sorting $10$ numbers. Some sorting algorithms run faster
if the input data is partially sorted already.  To minimize the
dependency on the specific instance we consider the {\em worst case
  time complexity} $T(n)$,
\begin{equation}
  \label{eq:def-worst-case-time}
  T(n) = \max_{|x|=n} t(x)
\end{equation}
where $t(x)$ is the running time of the algorithm for input data $x$
(in arbitrary units) and the maximum is taken over all problem
instances of size $n$. The worst case time is an upper bound for the
observable running time.

A measure of time complexity should be based on a unit of time that is
independent of the clock rate of a specific CPU. Such a unit is
provided by the time it takes to perform an elementary operation like
the addition of two integer numbers. Measuring the time in this unit
means counting the number of elementary operations executed by your
algorithm. This number in turn depends strongly on the implementation
details of the algorithm -- smart programmers and optimizing compilers
will try to reduce it.  Therefore we will not consider the precise
number $T(n)$ of elementary operations but only the asymptotic
behavior of $T(n)$ for large values of $n$ as denoted by the Landau
symbols $\bigo$ and $\Theta$:
\begin{itemize}
\item We say $T(n)$ is of order at most $g(n)$ and write
  $T(n)=\bigo(g(n))$ if there exist positive constants $c$ and $n_0$
  such that $T(n) \leq c g(n)$ for all $n \geq n_0$.
\item We say $T(n)$ is of order $g(n)$ and write $T(n)=\Theta(g(n))$
  if there exist positive constants $c_1,c_2$ and $n_0$ such that
  $c_1g(n) \leq T(n) \leq c_2 g(n)$ for all $n \geq n_0$.
\end{itemize}

Multiplying two $n\times n$ matrixes requires $n^3$ multiplications
according to the textbook formula.  Does this mean that the problem of
multiplying two $n\times n$ matrices has complexity $\Theta(n^3)$? No.
The textbook formula is a particular algorithm, and the time complexity
of an algorithm is only an {\em upper bound} for the inherent
complexity of a problem. In fact faster matrix multiplication
algorithms with complexity $\bigo(n^{\alpha})$ and $\alpha < 3$ have
been found during the last decades, the current record being
$\alpha=2.376$ \cite{coppersmith:winograd:90}. Since the product
matrix has $n^2$ entries, $\alpha$ can not be smaller than 2, and it
is an open question whether this lower bound can be achieved by an
algorithm. A problem where the upper bound from algorithmic complexity
meets an inherent lower bound is sorting $n$ items.  Under the general
assumption that comparisons between pairs of items are the only source
of information about the items, it can be shown that $\Theta(n\log n)$
is a lower bound for the number of comparisons to sort $n$ items in
the worst case \cite[chap.~9]{clr:algorithms}.  This bound is met by
algorithms like heapsort or mergesort.

\subsubsection{Problem size}

Our measure of time complexity still depends on the somewhat ambiguous
notion of problem size.  In the matrix multiplication example we
tacitly took the number $n$ of rows of one input matrix as the
``natural'' measure of size. Using the number of elements $m=n^2$
instead will ``speed up'' the $\bigo(n^3)$ algorithm to
$\bigo(m^{3/2})$ without changing a single line of program code. You
see that an unambiguous definition of problem size is required to
compare algorithms. 

In computational complexity, all problems which can be solved by a
polynomial algorithm, i.e.\ an algorithm with time complexity
$\Theta(n^k)$ for some $k$, are lumped together and called {\em
  tractable}. Problems which can only be solved by algorithms with
non-polynomial running time like $\Theta(2^n)$ or $\Theta(n!)$ are
also lumped together and called {\em intractable}.  There are
practical as well as theoretical reasons for this rather coarse
classification. One of the theoretical advantages is that it does
not distinguish between the $\bigo(n^3)$--
and $\bigo(n^{3/2})$--algorithm form above, hence we can afford some sloppiness
and stick with our ambigous ``natural'' measure of problem size. 

\subsection{Tractable and intractable problems}

\label{sec:tractable}

\subsubsection{Polynomial vs.\ exponential growth}

From a practical point of view, an exponential algorithm means a hard
limit for the accessible problem size. Suppose that with your current
equipment and you can solve a problem of size $n$ just within the
schedule. If you algorithm has complexity $\Theta(2^n)$, a problem of
size $n+1$ will need twice the time, bringing you definitely out of
schedule.  The increase in time caused by an $\Theta(n)$ or
$\Theta(n^2)$ algorithm on the other hand is far less dramatic and can
easily be compensated by upgrading your hardware. You might object
that a $\Theta(n^{100})$ algorithm outperforms a $\Theta(2^n)$
algorithm only for problem sizes that will never occur in your
application. A polynomial algorithm for a problem usually goes hand in
hand with a mathematical {\em insight} into the problem which enables
you to find a polynomial algorithm with small degree, typically
$\Theta(n^k)$, $k=1,2$ or $3$. Polynomial algorithms with $k > 10$ are
rare and arise in rather esoteric problems.

\subsubsection{Tractable trees}

As an example consider the following problem from network design. You
have a business with several offices and you want to lease phone lines
to connect them up with each other.  The phone company charges
different amounts of money to connect different pairs of cities, and
your task is to select a set of lines that connects all your offices
with a minimum total cost.

\begin{figure}[htb]
  \begin{center}
    \includegraphics[width=0.8\columnwidth]{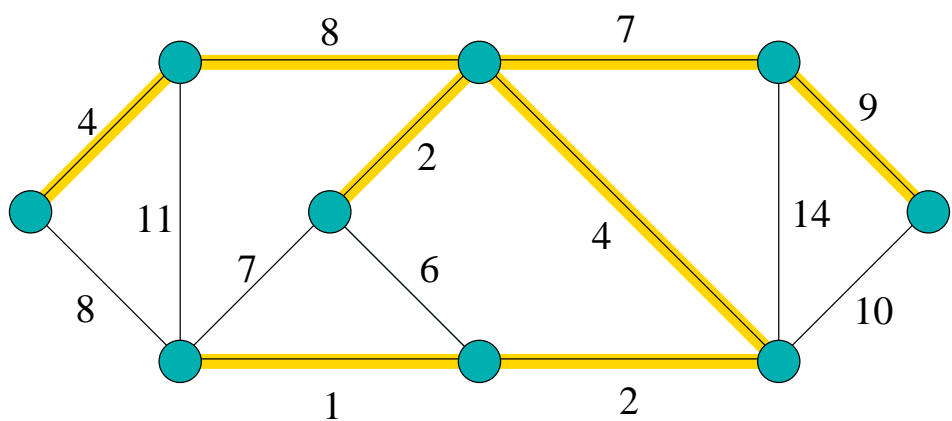}
  \end{center}
        \caption[Fig1]{\label{fig:mst}A weighted graph and its
          minimum spanning tree (colored edges).}
\end{figure}

In mathematical terms the cities and the lines between them form the
vertices $V$ and edges $E$ of a weigthed graph $G=(V,E)$, the weight
of an edge being the leasing costs of the corresponding phone line.
Your task is to find a subgraph that connects all vertices in the
graph, i.e.\ a spanning subgraph, and whose edges have minimum total
weight. Your subgraph should not contain cycles, since you can always
remove an edge from a cycle keeping all nodes connected and reducing
the cost. A graph without cycles is a tree, so what you are looking
for is a {\em minimum spanning tree} in a weighted graph
(Fig.~\ref{fig:mst}).  \probdef{Minimum Spanning Tree (MST)}{Given a
  weighted graph $G=(V,E)$ with non-negative weights. Find a spanning
  Tree $T\subseteq G$ with minimum total weight.}  How to find a minimum
spanning tree? A naive approach is to generate all possible trees with
$n$ vertices and keep the one with minimal weight. The enumeration of
all trees can be done using Pr\"ufer codes \cite{bollobas:98}, but
Cayley's formula tells us that there are $n^{n-2}$ different trees
with $n$ vertices.  Already for $n = 100$ there are more trees than
atoms in the observable universe!  Hence exhaustive enumeration is
prohibitive for all but the smallest trees.  The mathematical insight
that turns \problem{MST} into a tractable problem is
this:
\begin{quote}
  {\bf Lemma}: Let $U\subset V$ be any subset of the vertices of
  $G=(V,E)$, and let $e$ be the edge with the smallest weight of all
  edges connecting $U$ and $V-U$.  Then $e$ is part of the minimum
  spanning tree.
\end{quote}
{\em Proof}: By contradiction. Suppose $T$ is a minimum spanning tree
not containing $e$. Let $e=(u,v)$ with $u\in U$ and $v\in V-U$. Then
because $T$ is a spanning tree it contains a unique path from $u$ to
$v$, which together with $e$ forms a cycle in $G$. This path has to
include another edge $f$ connecting $U$ and $V-U$.  Now $T+e-f$ is
another spanning tree with less total weight than $T$. So $T$ was not
a minimum spanning tree.

The lemma allows to grow a minimum spanning tree edge by edge, using
Prim's algorithm for example:
\begin{algorithm}
  \algname{Prim}{$G$} \alginout{weighted graph $G(V,E)$}{minimum
    spanning tree $T\subseteq G$}
  \begin{algtab*}
    \textbf{begin}\\\algbegin
    Let $T$ be a single vertex $v$ from $G$\\
    \algwhile{$T$ has less than $n$ vertices}
    find the minimum edge connecting $T$ to $G-T$ \\
    add it to $T$ \\
    \algend \textbf{end}\\\algend \textbf{end}
  \end{algtab*}
\end{algorithm}     
The precise time complexity of Prim`s algorithm depends on the data
structure used to organize the edges, but in any case $\bigo(n^2\log
n)$ is an upper bound.  (see \cite{gabow:etal:86} for faster
algorithms).  Equipped with such a polynomial algorithm you can find
minimum spanning trees with thousands of nodes within seconds on a
personal computer. Compare this to exhaustive search! According to our
definition, \problem{mst} is a tractable problem.

\subsubsection{Intractable itineraries}

Encouraged by the efficient algorithm for \problem{mst}
we will now investigate a similar problem. Your task is to
plan an itinerary for a travelling salesman who must visit $n$ cities.
You are given a map with all cities and the distances between them. In
what order should the salesman visit the cities to minimize the total
distance he has to travel?  You number the cities arbitrarely and
write down the distance matrix $(d_{ij})$, where $d_{ij}$ denotes the
distance between city number $i$ and city number $j$. 
A tour is given by a {\em cyclic permutation} $\pi : [1\ldots n]\mapsto[1\ldots
  n]$, where $\pi(i)$ denotes the successor of city $i$, and your
problem can be defined as: \probdef{Traveling Salesman Problem
  (TSP)}{Given a $n\times n$ distance matrix with elements $d_{ij}\geq0$.
  Find a cyclic permutation $\pi : [1\ldots n]\mapsto[1\ldots
  n]$ that minimizes
\begin{equation}
  \label{eq:tsp-cost}
  c_n(\pi) = \sum_{i=1}^n d_{i\pi(i)}
\end{equation}}
The \problem{TSP} probably is the most famous
optimization problem, and there exists a vast literature specially
devoted to it, see
\cite{tspbook,reinelt:94,groetschel:padberg:99,tsp-homepage:link} and
references therein. It is not very difficult to find good solutions,
even to large problems, but how can we find the {\em best} solution for a
given instance? There are $(n-1)!$ cyclic permutations, calculating
the length of a single tour can be done in time $\bigo(n)$, hence
exhaustive search has complexity $\bigo(n!)$. Again this approach is
limited to very small instances. Is there a mathematical insight that 
provides us with a shortcut to the optimum solution, like for
\problem{mst}? Nobody knows! Despite the efforts of many brilliant
people, no polynomial algorithm for the \problem{tsp} has been found.
There are some smart and efficient (i.e.\ polynomial) algorithms that find 
good solutions but do not guarantee to yield the optimum \cite{reinelt:94}.
According to our definition, the \problem{tsp} is intractable.

\begin{figure}[htb]
  \begin{center}
    \includegraphics[width=0.8\columnwidth]{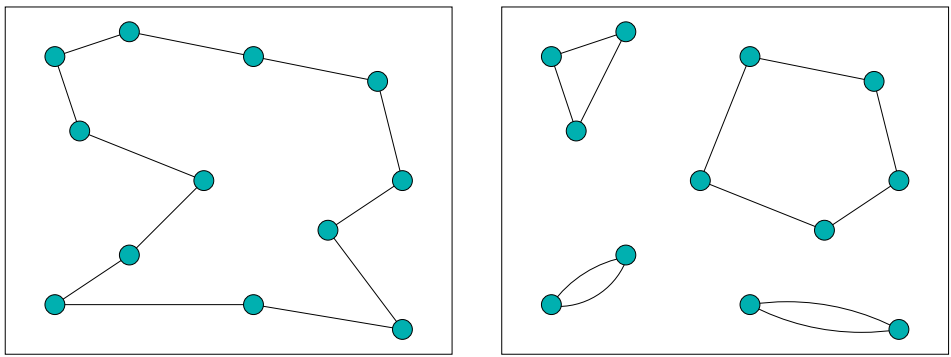}
  \end{center}
        \caption[Fig1]{\label{fig:tsp-ap}Same instance, different problems: A valid 
          configuration
          of the \problem{traveling salesman problem} (left) and the
          \problem{assignment} problem (right).  Whereas
          \problem{assignment} can be solved in polynomial time, the
          \problem{tsp} is intractable.}
\end{figure}

Why is the \problem{tsp} intractable? Again, nobody knows. There is no {\em proof} 
that excludes the existence of a polynomial algorithm for
\problem{tsp}, so maybe someday someone will
come up with a polynomial algorithm and the corresponding mathematical insight.
This is very unlikely, however, as we will see soon.

The intractability of the \problem{TSP} astonishes all the more considering the
tractability of a very similar, almost identical problem:
 \probdef{Assignment}
  {Given a $n\times n$ cost matrix with elements $d_{ij}\geq 0$.
  Find a permutation $\pi : [1\ldots n]\mapsto[1\ldots
  n]$ that minimizes
\begin{equation}
  \label{eq:ap-cost}
  c_n(\pi) = \sum_{i=1}^n d_{i\pi(i)}
\end{equation}}
The only difference between \problem{TSP} and \problem {Assignment} is
that the latter allows all permutations on $n$ items, not only the
cyclic ones.  If the $d_{ij}$ denote distances between cities,
\problem{Assignment} corresponds to total tour length minimization for
a variable number of salesmen, each travelling his own {\em subtour}
(Fig.\ \ref{fig:tsp-ap}).

The classical application of \problem{assignment} is the assignment of
$n$ tasks to $n$ workers, subject to the constraint that each worker
is assigned exactly one task. Let $d_{ij}$ denote the cost of having
task $j$ performed by worker $i$, and $\pi(i)$ denote the task
assigned to worker $i$, \problem{assignment} is the problem of
minimizing the total cost.

There are $n!$ possible assignments of $n$ tasks to $n$ workers, hence
exhaustive enumeration again is prohibitive. In contrast to the
\problem{TSP}, however, \problem{Assignment} can be solved in
polynomial time, for example using the $\bigo(n^3)$ hungarian
algorithm \cite{papadimitriou:steiglitz:82}.  Compared to
\problem{mst}, the algorithm and the underlying mathematical insight
are a bit more involved and will not be discussed here.

%%% Local Variables: 
%%% mode: latex
%%% TeX-master: t
%%% End: 

\section{Complexity classes}

\label{sec:classes}

\subsection{Decision problems}

\label{sec:decision}

So far we have discussed optimization problems: solving
\problem{MST}, \problem{TSP} or \problem{Assignment}
implies that we compare an exponential number of feasible solutions
with each other and pick the optimum.  Exhaustive search does this
explicitely, polynomial shortcuts implicitely.  Maybe we learn more
about the barrier that separates tractable from intractable problems
if we investigate simpler problems, problems where the solutions are
recognizable without explicit or implicit comparison to all feasible
solutions. So let us consider {\em decision problems}, problems whose
solution is either `yes' or `no'.

Any optimization problem can be turned into a decision problem by
adding a bound $B$ to the instance. Examples: \probdef{MST
  (Decision)}{Given a weighted graph $G=(V,E)$ with non-negative
  weights and a number $B\geq 0$. Does $G$ contain a spanning tree $T$
  with total weight $\leq B$?}  \probdef{TSP (Decision)} {Given a
  $n\times n$ distance matrix with elements $d_{ij}\geq0$ and a number
  $B\geq 0$.  Is there a tour $\pi$ with length $\leq B$?}  In a
decision problem, the feasible solutions are not evaluated relative to
each other but to an `absolut' criterion: a tour in the \problem{TSP}\ 
either has length $\leq B$ or not.

\problem{MST(D)} can be solved in polynomial time: simply solve the
optimization variant \problem{MST} and compare the result to the
parameter $B$. For the \problem{TSP(D)} this approach does not help.
In fact we will see in section \ref{sec:reductions} that there exists
a polynomial algorithm for \problem{TSP(D)} if and only if there
exists a polynomial algorithm for \problem{TSP}. It seems as if we
cannot learn more about the gap between tractable and intractable
problems by considering decision variants of optimization problems. So
lets look at other decision problems, not derived from optimization
problems.

\subsubsection{Eulerian circuits}
\label{sec:eulerian}

\begin{figure}[htb]
  \includegraphics[width=\columnwidth]{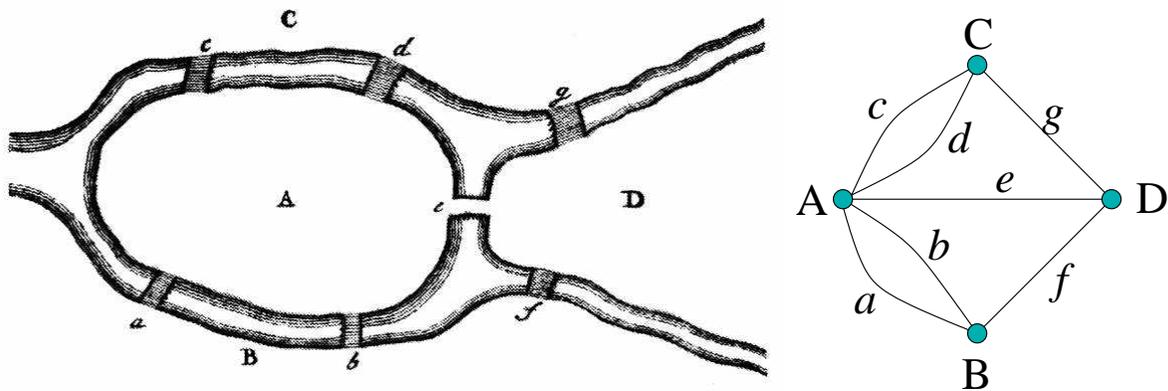}
        \caption[Fig1]{\label{fig:seven-bridges}The seven bridges of 
          K\"onigsberg,
          as drawn in Euler's paper from 1736 \cite{euler:1736} (left)
          and represented as a graph (right).  In the graph, the
          riverbanks and islands are condensed to points (vertices),
          and each of the bridges is drawn as a line (edge).}
\end{figure}

Our first `genuine' decision problem dates back into the 18th-century,
where in the city of K\"onigsberg (now Kaliningrad) seven bridges
crossed the river Pregel and its two arms
(Fig.~\ref{fig:seven-bridges}). A popular puzzle of the time asked if
it was possible to walk through the city crossing each of the bridges
exactly once and returning home.

It was Leonhard Euler who solved this puzzle in 1736
\cite{euler:1736}. First of all Euler recognizes that for the solution
of the problem the only thing that matters is the pattern of
interconnections of the banks and islands--a graph $G=(V,E)$ in modern
terms.  The graph corresponding to the puzzle of the K\"onigsberg
bridges has 4 vertices for the two banks and the two islands and 7
edges for the bridges (Fig.~\ref{fig:seven-bridges}). Euler's paper on
the K\"onigsberg bridges can be regarded as the birth of graph theory.

To generalize the K\"onigsberg bridges problem we need some
terminology from graph theory \cite{bollobas:98}.  A {\em walk} in a
graph $G=(V,E)$ is an alternating sequence of vertices $v\in V$ and
edges $(v,v')\in E$,
\begin{displaymath}
  v_1,(v_1,v_2),v_2,(v_2,v_3),\ldots,(v_{l-1},v_l),v_l.
\end{displaymath}
Note that the sequence begins and ends with a vertex, and each edge is
incident with the vertices immediately preceding and succeeding it. A
walk is termed {\em closed} if $v_l=v_1$, {\em open} otherwise.  A
walk is called a {\em trail} if all its edges are distinct, and a
closed trail is called a {\em circuit}.  What the strollers in
K\"onigsberg tried to find was a circuit that contains all edges. To
the honour of Leonhard Euler such a circuit is called {\em Eulerian
  circuit}.  Equipped with all this terminology we are ready to define
the generalization of the K\"onigsberg bridges problem:
\probdef{Eulerian Circuit}{Given a graph $G=(V,E)$. Does $G$ contain
  an Eulerian circuit?}  Obviously this is a decision problem: the
answer is either `yes' or `no', and any circuit can be checked to be
Eulerian or not without resorting to all possible circuits.

Once again exhaustive search would solve this problem, but the
intractability of this approach was already noticed by Euler. More
than 200 years before the advent of computers he wrote ``The
particular problem of the seven bridges of K\"onigsberg could be
solved by carefully tabulating all possible paths, thereby
ascertaining by inspection which of them, if any, met the requirement.
This method of solution, however, is too tedious and too difficult
because of the large number of possible combinations, and in other
problems where many more bridges are involved it could not be used at
all.'' (cited from \cite{lewis:papadimitriou:78}).

Euler solved the K\"onigsberg bridges problem not by listing all
possible trails but by mathematical insight. He noticed that in a
circuit you must leave each vertex via an edge different from the edge
that has taken you there. In other words, the degree of the vertex (that
is the number of edges adjacent to the vertex) must be even. This is
obviously a necessary condition, but Euler proved that it is also
sufficient:
\begin{quote}
  {\em Theorem:} A connected graph $G=(V,E)$ contains an Eulerian
  circuit if and only if the degree of every vertex $v\in V$ is even.
\end{quote}
Euler's theorem allows us to devise a polynomial algorithm for
\problem{Eulerian Circuit}: Check the degree of every vertex in the
graph. If one vertex has an odd degree, return `no'. If all vertices
have been checked having even degree, return `yes'. The running time
of this algorithm depends on the encoding of the graph. If $G=(V,E)$
is encoded as a $|V|\times|V|$ adjacency matrix with entries
$a_{ij}=$number of edges connecting $v_i$ and $v_j$, the the running
time is $\bigo(|V|^2)$.  Thanks to Euler, \problem{Eulerian Circuit}
is a tractable problem.  The burghers of K\"onigsberg on the other
hand had to learn from Euler, that they would never find a walk
through their hometown crossing each of the seven bridges exactly
once.

\subsubsection{Hamiltonian cycles}
\label{sec:hamiltonian}

\begin{figure}[htb]
  \begin{center}
    \includegraphics[width=\columnwidth]{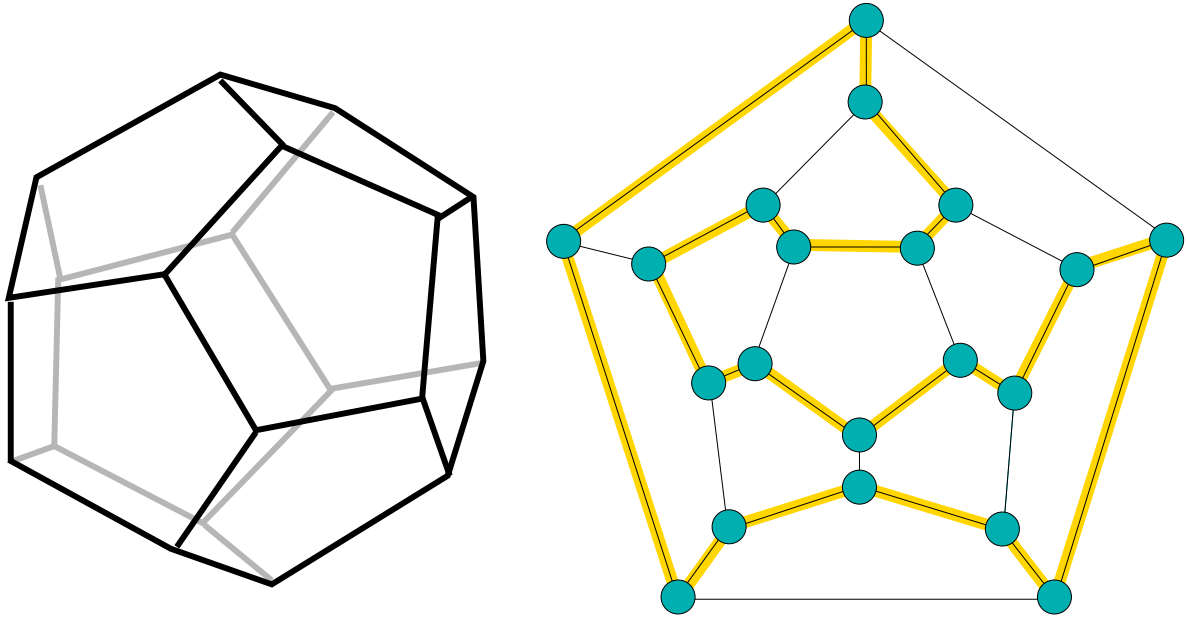}
  \end{center}
  \caption[Fig1]{\label{fig:icosian-game}Sir Hamilton's Icosian game: Find a route
    along the edges of of the dodecahedron (left), passing each corner
    exactly once and returning to the starting corner. A solution is
    indicated (shaded edges) in the planar graph that is isomorphic to
    the dodecahedron (right).}
\end{figure}

Another decision problem is associated with the mathematician and
Astronomer Royal of Ireland, Sir William Rowan Hamilton. In the year
1859, Hamilton put on the market a new puzzle called the Icosian game
(Fig.~\ref{fig:icosian-game}).

The generalization of the Icosian game calls for some definitions from
graph theory: A closed walk in a graph is called a {\em cycle} if all
its vertices (execpt the first and the last) are distinct. A
Hamiltonian cycle is a cycle that contains all vertices of a graph.
The generalization of the Icosian game then reads \probdef{Hamiltonian
  Cycle}{Given a graph $G=(V,E)$. Does $G$ contain a Hamiltonian
  cycle?}  There is a certain similarity between \problem{Eulerian
  Circuit} and \problem{Hamiltonian Cycle}.  In the former we must
pass each edge once; in the latter, each vertex once.  Despite this
resemblance the two problems represent entirely different degrees of
difficulty.  The available mathematical insights into
\problem{Hamiltonian Cycle} provide us neither with a polynomial
algorithm nor with a proof that such an algorithm is impossible.
\problem{Hamiltonian Cycle} is intractable, and nobody knows why.

\subsubsection{Coloring}
\label{sec:coloring}

Imagine we wish to arrange talks in a congress in such a way that no
participant will be forced to miss a talk he would like to hear.
Assuming a good supply of lecture rooms enabling us to hold as many
parallel talks as we like, can we finish the programme within $k$ time
slots?  This problem can be formulated in terms of graphs: Let $G$ be
a graph whose vertices are the talks and in which two talks are
adjacent (joined by an edge) if and only if there is a participant
whishing to attend both. Your task is to assign one of the $k$ time
slots to each vertex in such a way that adjacent vertices have
different time slots. The common formulation of this problem uses
colors instead of time slots: \probdef{$k$-Coloring}{Given a graph
  $G=(V,E)$. Is there a coloring of the vertices of $G$ using at most
  $k$ different colors such that no two adjacent vertices have the
  same color?}  When $k=2$ this problem is tractable --- the
construction of a polynomial algorithm is left as easy exercise.  For
$k=3$ things change considerably: \problem{$3$-Coloring} is intractable.
Note that for larger $k$ the problem gets easier again: a {\em planar}
graph is always colorable with $4$ colors! This is the famous $4$-color
Theorem. \problem{$3$-Coloring} remains intractable even when restricted
to planar graphs.

\subsubsection{Satisfiability}
\label{sec:sat}

We close this section with a decision problem that is not from graph
theory but from Boolean logic. A Boolean variable $x$ can take on the
value 0 (false) or 1 (true). Boolean variables can be combined in {\em
  clauses} using the Boolean operators
\begin{itemize}
\item[--] NOT $\overline{\cdot}$ (negation): the clause $\overline{x}$
  is true ($\overline{x}=1$) if and only if $x$ is false ($x=0$).
\item[--] AND $\wedge$ (conjunction): the clause $x_1\wedge x_2$ is
  true ($x_1\wedge x_2 = 1$) if and only if both variables are true:
  $x_1=1$ and $x_2=1$
\item[--] OR $\vee$ (disjunction): the clause $x_1\vee x_2$ is true
  ($x_1\vee x_2 = 1$) if and only if at least one of the variables is
  true: $x_1=1$ or $x_2=1$.
\end{itemize}
A variable $x$ or its negation $\overline{x}$ is called a {\em
  literal}.  Different clauses can be combined to yield complex
Boolean formulas, e.g.
\begin{equation}
  \label{eq:sat-example}
  F_1(x_1,x_2,x_3) = (x_1\vee\overline{x_2}\vee x_3) \wedge (x_2\vee\overline{x_3}) \wedge
                   (\overline{x_1}\vee x_2) \wedge (\overline{x_1}\vee\overline{x_3}).
\end{equation}
A Boolean formula evaluates to either 1 or 0, depending on the
assignment of the Boolean variables. In the example above $F_1=1$ for
$x_1=1, x_2=1, x_3=0$ and $F_1=0$ for $x_1=x_2=x_3=1$.  A formula $F$
is called {\em satisfiable}, if there is at least one assignment of
the variables such that the formula is true. $F_1$ is satisfiable,
\begin{equation}
  \label{eq:unsat-example}
  F_2(x_1,x_2) = (\overline{x_1}\vee x_2) \wedge \overline{x_2} \wedge x_1
\end{equation}
is not satisfiable.

Every Boolean formula can be written in {\em conjunctive normal form}
(CNF) i.e.\ as a set of clauses $C_k$ combined exclusively with the
AND--operator
\begin{equation}
  \label{eq:def-CNF}
  F = C_1 \wedge C_2 \wedge \cdots \wedge C_m
\end{equation}
where the literals in each clause are combined exclusively with the
OR--operator.  The examples $F_1$ and $F_2$ are both written in CNF.
Each clause can be considered as a constraint on the variables, and
satisfying a formula means satisfying a set of (possibly conficting)
constraints simultaneously. Hence \probdef{Satisfiability (SAT)}{
  Given disjunctive clauses $C_1,C_2,\ldots,C_m$ of literals, where a
  literal is a variable or negated variable from the set
  $\{x_1,x_2,\ldots,x_n\}$. Is there an assignment of variables that
  satisfies all clauses simultaneously?}  can be considered as
prototype of a constraint satisfaction problem \cite{kumar:92}.
Fixing the number of literals in each clause leads to
\probdef{$k$-SAT}{ Given disjunctive clauses $C_1,C_2,\ldots,C_m$ of
  $k$ literals each, where a literal is a variable or negated variable
  from the set $\{x_1,x_2,\ldots,x_n\}$. Is there an assignment of
  variables that satisfies all clauses simultaneously?  } Polynomial
algorithms are known for \problem{$1$-SAT} and \problem{$2$-SAT}
\cite{aspvall:plass:tarjan:79}.  No polynomial algorithm is known for
general \problem{SAT} and \problem{$k$-SAT} if $k > 2$.

\subsection{The classes \p\ and \np}

\label{sec:p-and-np}

\subsubsection{Tractable problems}

Now we have seen enough examples to introduce two important complexity
classes for decision problems.

The class of tractable decision problems is easy to define: it
consists of those problems, for which a polynomial algorithm is known.
The corresponding class is named \p\ for ``polynomial'':
\begin{quote}
  {\bf Definition:} A decision problem $P$ is element of the class \p\ 
  if and only if it can be solved by a polynomial time algorithm.
\end{quote}
\problem{Eulerian Circuit}, \problem{$2$-Coloring}, \problem{MST(D)}
etc.\ are in \p.

\subsubsection{Nondeterministic algorithms}

\begin{figure}[tb]
  \begin{center}
    \includegraphics[width=0.8\columnwidth]{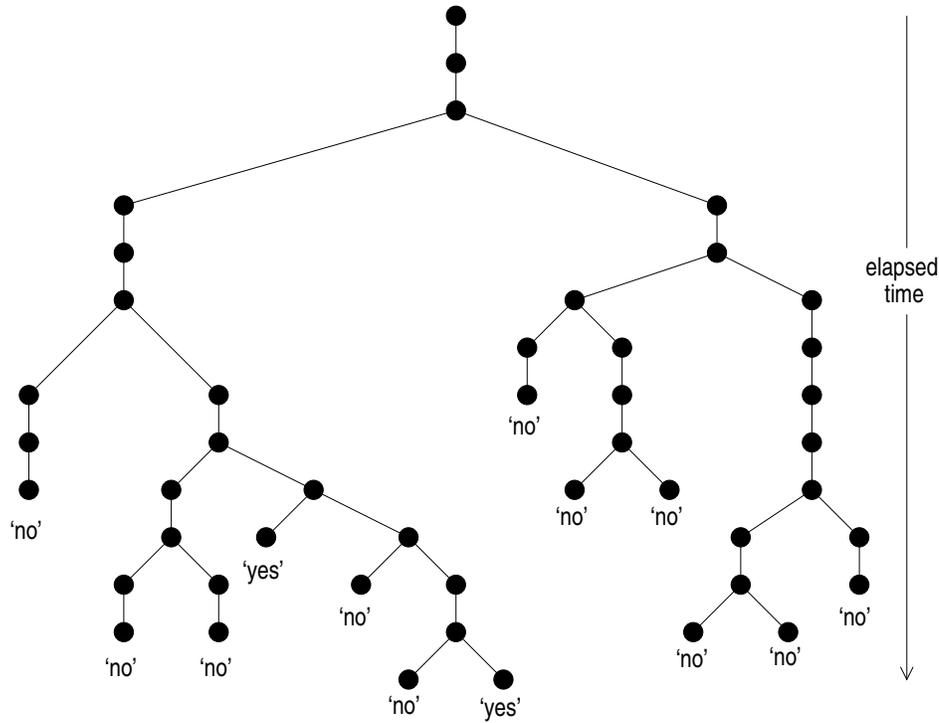}
  \end{center}
  \caption[Fig1]{\label{fig:np-alg}Example of the execution history of a nondeterministic
    algorithm.}
\end{figure}

The definition of the second complexity class involves the concept of
a {\em nondeterministic algorithm}. A nondeterministic algorithm is
like an ordinary algorithm, except that it may use one additional,
very powerful instruction \cite{johnson:papadimitriou:85}:
\begin{quote}
  {\bfseries goto both} label 1, label 2
\end{quote}
This instruction splits the computation into two parallel processes,
one continuing from each of the instructions indicated by ``label 1''
and ``label 2''. By encountering more and more such instructions, the
computation will branch like a tree into a number of parallel
computations that potentially can grow as an exponential function of
the time elapsed (see Fig.~\ref{fig:np-alg}). A nondeterministic
algorithm can perform an exponential number of computations in
polynomial time! In the world of conventional computers,
nondeterministic algorithms are a {\em theoretical concept} only, but
in quantum computing this may change
(section~\ref{sec:quantum}).  We need the concept of nondeterminism to
define the class \np\ of ``nondeterministic polynomial'' problems:
\begin{quote}
  {\bf Definition}: A decision problem $P$ is in the class \np, if and
  only if it can be solved in {\em polynomial} time by a {\em
    nondeterministic algorithm}.
\end{quote}
Solubility by a nondeterministic algorithm means this: All branches of
the computation will stop, returning either `yes' or `no'. We say that
the overall algorithm returns `yes', if {\em any} of its branches
returns `yes'.  The answer is `no', if none of the branches reports
`yes'. We say that a nondeterministic algorithm solves a decision
problem in polynomial time, if the number of steps used by the first
of the branches to report `yes' is bounded by a polynomial in the size
of the problem.

We require polynomial solubility only for the `yes'--instances of
decision problem.  This asymmetry between `yes' and `no' reflects the
asymmetrie between the `there is'-- and `for all'--quantifiers in
decision problems: a graph $G$ is a `yes'--instance of
\problem{Hamiltonian Cycle}, if {\em there is} at least one
Hamiltonian cycle in $G$. For a `no'--instance, {\em all} cycles in
$G$ have to be non Hamiltonian.

Note that the conventional (deterministic) algorithms are special
cases of a nondeterministic algorithms (those nondeterministic
algorithms that do not use the {\bf goto both} instruction). It
follows immediately that \p$\subseteq$\np.

All decision problems we have discussed in the preceeding section are
members of \np.  Here's a nondeterministic polynomial algorithm for
\problem{SAT}:
\begin{algorithm}
  \algname{Satisfiability}{$F$} \alginout{Boolean formula
    $F(x_1,\ldots,x_n)$}{'yes' if $F$ is satisfiable, 'no' otherwise}
  \begin{algtab*}
    \textbf{begin}\\\algbegin \algforto{$i=1$}{n}
    \textbf{goto both} label 1, label 2 \\
    label 1: $x_i = $ true; \textbf{continue} \\
    label 2: $x_i = $ false; \textbf{continue} \\
    \algend \textbf{end}\\
    \algifthenelse{$F(x_1,\ldots,x_n)=$ true}{\algreturn
      `yes'}{\algreturn `no'} \algend \textbf{end}
  \end{algtab*}
\end{algorithm}
The \textbf{for}--loop branches at each iteration: in one branch
$x_i=$ true, in the other branch $x_i=$ false (the \textbf{continue}
instruction starts the next iteration of the loop). After executing
the \textbf{for}--loop we have $2^n$ branches of computation, one for
each of the possible assignments of $n$ Boolean variables.

The power of nondeterministic algorithms is that they allow the
exhaustive enumeration of an exponentially large number of candidate
solutions in polynomial time.  If the evaluation of each candidate
solution (calculating $F(x_1,\ldots,x_n)$ in the above example) in
turn can be done in polynomial time, the total nondeterministic
algorithm is polynomial.  For a problem from the class \np, the sole
source of intractability is the exponential size of the search space.

\subsubsection{Succinct certificates}

There is a second, equivalent definition of \np, based on the notion
of a {\em succinct certificate}.  A certificate is a proof. If you
claim that a graph $G$ has a Hamiltonian cycle, you can proof your
claim by providing a Hamiltonian cycle. Certificates for
\problem{Eulerian Circuit} and \problem{$k$-Coloring} are an Eulerian
circuit and a valid coloring. A certificate is succinct, if its
size is bounded by a polynomial in the size of the problem. The second
definition then reads
\begin{quote}
  {\bf Definition:} A decision problem $P$ is element of the class
  \np\ if and only if for every `yes'--instance of $P$ there exists a
  {\em succinct certificate} that can be verified in polynomial time.
\end{quote}
The equivalence of both definitions can easily be shown
\cite{johnson:papadimitriou:85}. The idea is that a succinct
certificate can be used to deterministically select the branch in a
nondeterministic algorithm that leads to a `yes'-output.

The definition based on nondeterministic algorithms reveals the key
feature of the class \np\ more clearly, but the second definition is
more usefull for proving that a decision problem is in \np. As an
example consider \probdef{Compositeness}{Given a positive integer $N$.
  Are there integer numbers $p>1$ and $q>1$ such that $N=pq$?}  A
certificate of a `yes' instance $N$ of \problem{Compositeness} is a
factorization $N=pq$. It is succinct, because the number of bits in
$p$ and $q$ is less or equal the number of bits in $N$, and it can be
verified in quadratic time by multiplication. Hence
\problem{Compositeness}$\in$\np.

Most decision problems ask for the existence of an object with a given
property, like a cycle which is Hamiltonian or a factorization with
integer factors.  In these cases, the desired object may serve as a
succinct certificate.  For some problems this does not work, however,
like for \probdef{Primality}{Given a positive integer $N$. Is $N$
  prime?}  \problem{Primality} is the negation or complement of
\problem{Compositeness}: the `yes'--instances of the former are the
`no'--instances of the latter and vice versa. A succinct certificate
for \problem{Primality} is by no means obvious.  In fact, for many
decision problems in \np\ no succinct certificate is kown for the
complement, i.e.\ it is not known whether the complement is also in
\np. For \problem{Primality} however, one can construct a succinct
certificate based on Fermat's Theorem \cite{pratt:75}.  Hence
\problem{Primality}$\in$\np.

\subsubsection{A first map of complexity}

\begin{figure}[htb]
  \begin{center}
    \includegraphics[width=0.7\columnwidth]{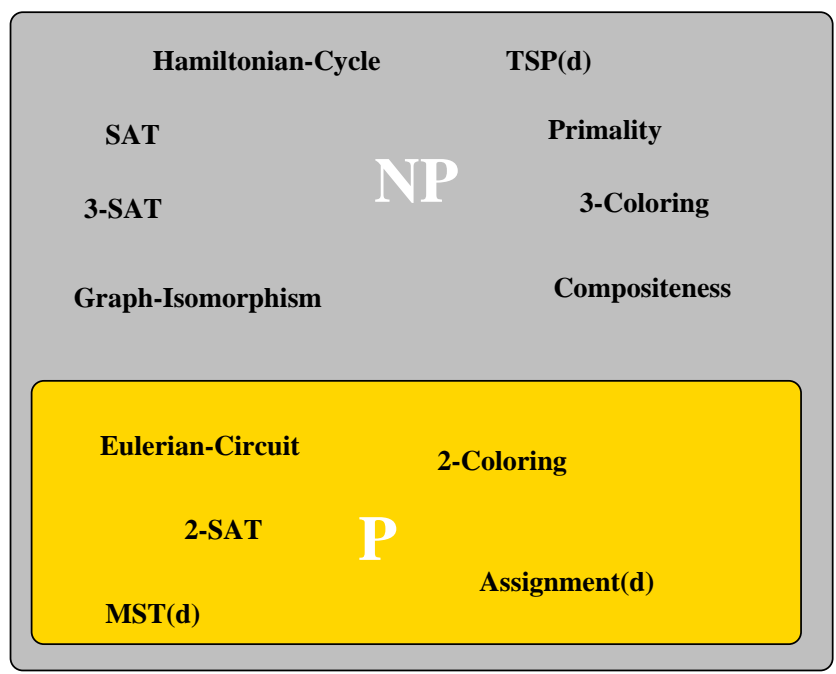}
  \end{center}
  \caption[Fig1]{\label{fig:np-map-1}A first map of complexity. All problems 
    indicated are defined within the text. Problems with a (D) are decision variants
    of optimization problems.  }
\end{figure}

Fig.~\ref{fig:np-map-1} summarizes what we have achieved so far. The
class \np\ consists of all decision problems whose sole source of
difficulty is the size of the search space which grows exponentially
with the size of the problem. These problems are intractable unless
some mathematical insight provides us with a polynomial shortcut to
avoid exhaustive search.  Such an insight promotes a problem into the
class \p\ of polynomially soluble problems.

The class \np\ not only contains a large number of problems with
important applications, but additionally represents a real challenge:
all problems in \np\ still have a chance to be in \p.  A proof of
non-existence of a polynomial algorithm for a single problem from \np\ 
would establish that $\mathcal{P}\neq\mathcal{NP}$. As long as such a
proof is missing,
\begin{equation}
  \label{eq:p-np-conjecture}
  \mathcal{P} \stackrel{?}{=} \mathcal{NP}
\end{equation}
represents the most famous open conjecture in theoretical computer
science.

\subsection{\np-completeness}

\subsubsection{Polynomial reductions}
\label{sec:reductions}

The computational complexities of two
problems $P_1$ and $P_2$ can be related to each other by constructing
an algorithm for $P_1$ that uses an algorithm for $P_2$ as a
``subroutine''.  Consider the following algorithm that relates
\problem{Hamiltonian Cycle} to \problem{TSP(D)}:
\begin{algorithm}
  \algname{Hamiltonian Cycle}{$G$} \alginout{Graph $G=(V,E)$} {'yes'
    if $G$ contains a Hamiltonian cycle, 'no' otherwise}
  \begin{algtab}
    \textbf{begin}\\\algbegin
    $n := |V|$\\
    \algforto{$i=1$}{n} \algforto{$j=1$}{n}
    \algifthenelse{$(v_i,v_j)\in E$}{$d_{ij} := 1$}{$d_{ij} := 2$}
    \algend \algend \algifthenelse{TSP-decision ($d, B:=n$) =
      `yes'}{\algreturn `yes'}{\algreturn `no'} \algend \textbf{end}
  \end{algtab}
\end{algorithm}
This algorithm solves \problem{Hamiltonian Cycle} by solving an
appropriate instance of \problem{TSP(D)}. In the \textbf{for}--loops
(lines 3-5) a distance matrix $d$ is set up with entries $d_{ij}=1$ if
there is an edge $(v_i,v_j)$ in $G$ and $d_{ij}=2$ otherwise. Now a
Hamiltonian cycle in $G$ is valid tour in the corresponding
\problem{TSP} with all intercity distances having length $1$, i.e.\ 
with total length $n$.  Conversely, suppose that the \problem{TSP} has
a tour of length $n$. Since the intercity distances are either $1$ or
$2$ and a tour sums up $n$ such distances, a total length $n$ implies
that each pair of successively visited cities must have distance $1$,
i.e.\ the tour follows existing edges in $G$ and corresponds to a
Hamiltonian cycle. Hence the call to a subroutine that solves
\problem{TSP(D)} (line 7) yields a solution to \problem{Hamiltonian
  Cycle}.

How does this algorithm relate the computational complexity of
\problem{Hamiltonian Cycle} to that of \problem{TSP(D)}? Note that
this is a polynomial algorithm if the call to the
\problem{TSP(D)}--solver is considered as elementary operation. If
someone comes up with a polynomial algorithm for \problem{TSP(D)}, we
will instantly have a polynomial algorithm for \problem{Hamiltonian
  Cycle}! We say that \problem{Hamiltonian Cycle} is {\em polynomially
  reducible} to \problem{TSP(D)} and write
\begin{equation}
  \label{eq:hc-tsp}
  \mbox{\problem{Hamiltonian Cycle}} \leq \mbox{\problem{TSP(D)}}.
\end{equation}
In many books, polynomial reducibility is denoted by `$\propto$'
instead of `$\leq$'.  We follow \cite{ausiello:etal:99} and use
`$\leq$' because this notation stresses an important consequence of
polynomial reducibility: the existence of a polynomial reduction from
$P_1$ to $P_2$ excludes the possibility that $P_2$ can be solved in
polynomial time, but $P_1$ cannot. Hence $P_1\leq P_2$ can be read as
$P_1$ {\em cannot be harder than} $P_2$. Here is the (informal)
definition:
\begin{quote}
  {\bf Definition:} We say a problem $P_1$ is {\em polynomially
    reducible} to a problem $P_2$ and write $P_1\leq P_2$ if there
  exists a polynomial algorithm for $P_1$ provided there is a
  polynomial algorithm for $P_2$.
\end{quote}

\subsubsection{\np-complete problems}

Here are some other polynomial reductions that can be verified similar to
Eq.~\ref{eq:hc-tsp}:
\begin{eqnarray}
  \label{eq:reductions}
  \mbox{\problem{SAT}} &\leq& \mbox{\problem{$3$-SAT}}\nonumber\\
  \mbox{\problem{$3$-SAT}} &\leq& \mbox{\problem{3-Coloring}}\\
  \mbox{\problem{3-Coloring}} &\leq& \mbox{\problem{Hamiltonian Cycle}}\nonumber
\end{eqnarray}
See \cite{garey:johnson:79,karp:72} for the corresponding reduction
algorithms.  Polynomial reducibility is transitive: $P_1\leq P_2$ and
$P_2\leq P_3$ imply $P_1\leq P_3$. From transitivity and
Eqs.~\ref{eq:hc-tsp} and \ref{eq:reductions} it follows that each of
\problem{SAT}, \problem{$3$-SAT}, \problem{$3$-Coloring} and
\problem{Hamiltonian Cycle} reduces to \problem{TSP(D)}, i.e.\ a
polynomial algorithm for \problem{TSP(D)} implies a polynomial
algorithm for all these problems! This is amazing, but only the
beginning. The true scope of polynomial reducibility was revealed by
Stephen Cook in 1971 \cite{cook:71} who proved the following,
remarkable theorem:
\begin{quote}
  {\bf Theorem:} (Cook, 1971) All problems in \np\ are polynomially
  reducible to \problem{SAT},
  \begin{equation}
    \label{eq:cook}
    \forall P\in\mathcal{NP}: P \leq \mbox{\problem{sat}}
  \end{equation}
\end{quote}
This theorem means that
\begin{enumerate}
\item no problem in \np\ is harder than \problem{SAT} or \problem{SAT}
  is among the hardest problems in \np.
\item a polynomial algorithm for \problem{SAT} would imply a
  polynomial algorithm for {\em every} problem in \np, i.e.\ would
  imply $\mathcal{P} = \mathcal{NP}$.
\end{enumerate}
It seems as if \problem{SAT} is very special, but according to
transitivity and Eqs.~\ref{eq:reductions} and \ref{eq:hc-tsp} it can
be replaced by \problem{$3$-SAT}, \problem{$3$-Coloring},
\problem{Hamiltonian Cycle} or \problem{TSP(D)}. These problems form a
new complexity class:
\begin{quote}
  {\bf Definition:} A problem $P$ is called \np-complete if
  $P\in\mathcal{NP}$ and $Q \leq P$ for all $Q\in\mathcal{NP}$
\end{quote}
The class of \np-complete problems collects the hardest problems in
\np.  If any of them has an efficient algorithm, then {\em every}
problem in \np\ can be solved efficiently, i.e.\ $\mathcal{P} =
\mathcal{NP}$. This is extremely unlikely, however, considered the
futile efforts of many brilliant people to find polynomial algorithms
for problems like \problem{Hamiltonian Cycle} or \problem{TSP(D)}.

\begin{figure}[htb]
  \begin{center}
    \includegraphics[width=0.7\columnwidth]{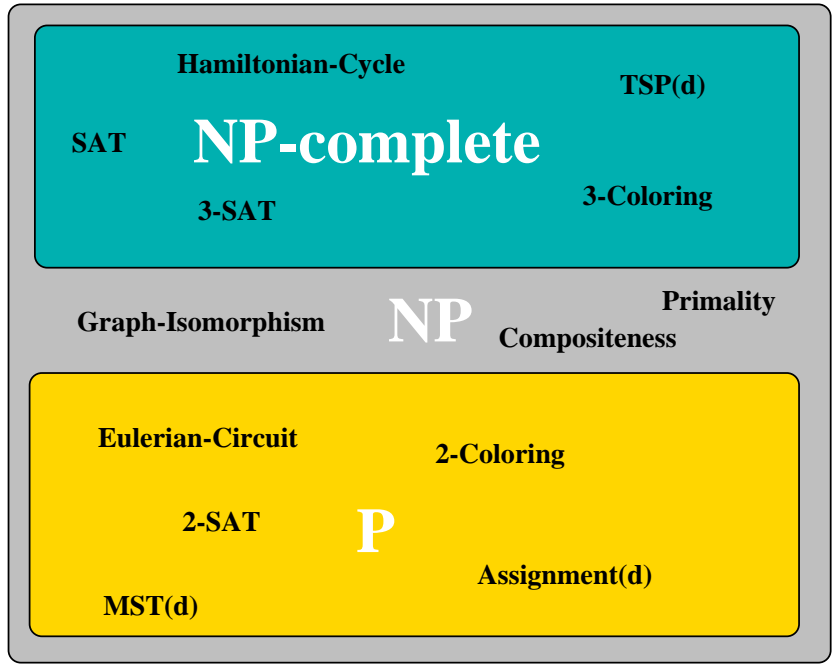}
  \end{center}
  \caption[Fig1]{\label{fig:np-map-2}The map of complexity revisited. 
    }
\end{figure}

\subsubsection{The map of \np }

Since Cook's Theorem many problems have been shown to be \np-complete.
A comprehensive, up-to-date list of hundreds of \np-complete problems
can be found in the web \cite{np-compendium:link}.  Our map of \np\ 
needs some redesign (Fig.~\ref{fig:np-map-2}). It turns out that all
the intractable problems we have discussed so far are \np-complete --
except \problem{Compositeness} and \problem{Primality}. For both
problems neither a polynomial algorithm is known nor a polynomial
reduction that clasifies them \np-complete. Another \np\ problem which
resists classification in either \p\ or \np\ is this: \probdef{Graph
  Isomorphism}{Given two graphs $G=(V,E)$ and $G´(V,E´)$ on the same
  set of nodes.  Are $G$ and $G´$ isomorphic, i.e.\ is there a
  permutation $\pi$ of $V$ such that $G´=\pi(G)$, where by $\pi(G)$ we
  denote the graph $(V,\{[\pi(u),\pi(v)] : [u,v]\in E\})$?}

\begin{figure}[htb]
  \includegraphics[width=\columnwidth]{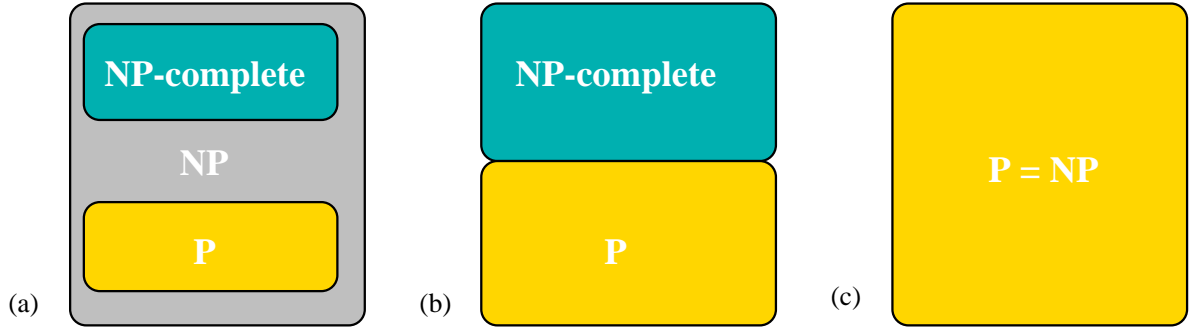}
        \caption[Fig1]{\label{fig:np-scenarios}
          Three tentative maps of \np. (b) can be ruled out. It is
          very likely (but not sure) that (a) is the correct map.  }
\end{figure}

\noindent
There are more problems in \np\ that resists a classification in \p\ 
or \np, but none of these problems has been {\em proven} not to belong
to \p\ or \np. What has been proven is
\begin{quote}
  {\bf Theorem:} If $\mathcal{P}\neq\mathcal{NP}$, then there exist
  \np\ problems which are neither in \p\ nor are they \np-complete.
\end{quote}
This Theorem \cite{ladner:75} rules out one of three tentative maps of
\np\ (Fig.~\ref{fig:np-scenarios}).

\subsection{Beyond \np}

\subsubsection{Optimization problems}
\label{sec:optimization}

How does the classification of decision problems relates to
optimization problems?  The general instance of an optimization
problem is a pair $(F,c)$, where $F$ is the set of feasible solutions
and $c$ is a cost function $c: F\to \mathbb{R}$. We will consider only
{\em combinatorial optimization} where the set $F$ is countable.  A
combinatorial optimization problem $P$ comes in three different
flavors:
\begin{enumerate}
\item The optimization problem $P(O)$: Find the {\em feasible
    solution} $f^*\in F$ that minimizes the cost function.
\item The evaluation problem $P(E)$: Find the {\em cost} $c^* =
  c(f^*)$ of the minimum solution.
\item The decision problem $P(D)$: Given a bound $B\in\mathbb{R}$, is
  there a feasible solution $f\in F$ such that $c(f) \leq B$?
\end{enumerate}
Under the assumption that the cost function $c$ can be evaluated in
polynomial time, it is straightforward to write down polynomial
reductions that establish
\begin{equation}
  \label{eq:d-e-o}
  P(D) \leq P(E) \leq P(O).
\end{equation}
If the decision variant of an optimization problem is \np-complete,
there is no efficient algorithm for the optimization problem at
all---unless $\mathcal{P}=\mathcal{NP}$.  An optimization problem like
\problem{TSP}, whose decision variant is \np-complete is denoted
\np-hard.

Does a polynomial algorithm for a decison problem imply a polynomial
algorithm for the optimization or evaluation variant?  For that we
need to proof the reversal of Eq.~\ref{eq:d-e-o},
\begin{equation}
  \label{eq:o-e-d}
  P(O) \leq P(E) \leq P(D).
\end{equation}
$P(E)\leq P(D)$ can be shown to hold if the cost of the optimum
solution is an integer with logarithm bounded by a polynomial in the
size of the input. The corresponding polynomial reduction evaluates
the optimal cost $c^*$ by asking the question ``Is $c^* \leq B$?'' for
a sequence of values $B$ that approaches $c^*$, similar to the
bisection method to find the zeroes of a function.

There is no general method to prove $P(O) \leq P(E)$, but a strategy
that often works can be demonstrated for the \problem{TSP}: Let $c^*$
be the known solution of \problem{TSP(E)}.  Replace an arbitrary entry
$d_{ij}$ of the distance matrix with a value $c > c^*$ and solve
\problem{TSP(E)} with this modified distance matrix. If the length of
the optimum tour is not affected by this modification, the link $ij$
does not belong to the optimal tour. Repeating this procedure for
different links one can reconstruct the optimum tour with a polynomial
number of calls to a \problem{TSP(E)}--solver, hence
$\mbox{\problem{TSP(O)}}\leq\mbox{\problem{TSP(E)}}$.

\subsubsection{Counting problems}
\label{sec:counting}

So far we have studied two related styles of problems: Decision
problems ask whether a desired solution exists, optimization problems
require that a solution be produced. A third important and
fundamentally different kind of problem asks, how many solutions
exist.  The {\em counting} variant of \problem{SAT} reads
\probdef{\#SAT}{Given a Boolean expression, compute the number of
  different truth assignments that satisfy it} Similarly,
\problem{\#Hamiltonian Cycle} asks for the number of different
Hamiltonian cycles in a given graph, \problem{\#TSP} for the number of
different tours with length $\leq B$ and so on.
\begin{quote}
  {\bf Definition:} A counting problem $\mbox{\#}P$ is a pair $(F,d)$,
  where $F$ is the set of all feasible solutions and $d$ is a decision
  function $d: F\mapsto \{0,1\}$. The output of $\mbox{\#}P$ is the
  number of $f\in F$ with $d(f)=1$.  The class \#\p\ (pronounced
  ``number P'') consists of all counting problems associated with a
  decision function $d$ that can be evaluated in polynomial time.
\end{quote}
Like the class \np, \#\p\ collects all problems whose sole source of
intractability is the number of feasible solutions.  A polynomial
algorithm for a counting problem $\mbox{\#}P$ implies a polynomial
algorithm for the associated decision problem $P$: $P\leq \mbox{\#}P$.
Hence it is very unlikely that \problem{\#SAT} can be solved
efficiently. In fact one can define polynomial reducibility for
counting problems and prove that all problems in \#\p\ reduce
polynomially to \problem{\#SAT} \cite{papadimitriou:94}:
\begin{quote}
  {\bf Theorem:} \problem{\#SAT} is \#\p-complete.
\end{quote}
As you might have guessed, \problem{\#Hamiltonian Cycle} and
\problem{\#TSP} are \#\p-complete, too. Despite the similarity between
\np\ and \#\p, counting problems are inherently harder than decision
problems. This is documented by those \#\p-complete problems, for
which the corresponding decision problem can be solved in polynomial
time, the classical example being the problem of calculating the
permanent of a matrix \cite{valiant:79}.

%%% Local Variables: 
%%% mode: latex
%%% TeX-master: "lecture"
%%% End: 

\section{Computational complexity and physics}
\label{sec:physics}

\subsection{Quantum parallelism}
\label{sec:quantum}

In a seminal paper \cite{feynman:82}, Richard Feynman pointed out that
a system of $n$ quantum particles is exponentially hard to simulate on
a classical computer.  The idea of quantum computing is to reverse
this situation and simulate a classically hard (i.e.\ exponential!)
problem in polynomial time on a computer made of quantum devices.

A quantum computer processes qubits, quantum two-state systems
$|0\rangle$, $|1\rangle\rangle$.  A key feature of a quantum computer
is that its registers can hold and process linear superpositions of
all $2^n$ product states of $n$ qubits like
\begin{equation}
  \label{eq:qubit}
  \frac{1}{\sqrt{2^n}}\sum_{i_1,i_2,\ldots,i_n=0}^1 |i_1 i_2 \cdots i_n\rangle.
\end{equation}
Using this feature it is not very difficult to construct a quantum
computer capable of computing any function $f(x_1,\ldots,x_n)$ of $n$
Boolean variables simultaneously for all $2^n$ possible input values
--- in theory at least. This {\em quantum parallelism} resembles very
much a nondeterministic algorithm with its {\bf goto both} instruction
and its exponentially branching execution tree.  Is quantum
parallelism the key to exponential computing power? The problem is how
to extract the exponential information out of a quantum computer. When
we defined nondeterministic solubility we did not care about how to
``spot'' the single `yes'--answer among the $2^n$ `no'--answers. This
works fine for a theoretical concept, but for a practical computer
reading the output matters a lot.

In order to gain advantage of exponential parallelism, it needs to be
combined with another quantum feature known as interference. The goal
is to arrange the computation such that constructive interference
amplifies the result we are after and destructive interference cancels
the rest. Due to the importance of interference phenomena it is not
surprising that calculating the Fourier transform was the first
problem that undergoes an exponential speedup: from $\bigo(n\log n)$
on a classical to $\bigo(\log^2n)$ on a quantum computer. This speedup
was the seed for the most important quantum algorithm known today:
Shor's algorithm to factor an integer in polynomial time
\cite{shor:97}.

Although Shor's algorithm has some consequences for public key cryptography,
it does not shatter the world of \np: remember that \problem{Compositeness}
is in \np, but not \np-complete. Hence the holy grail of quantum computing,
a polynomial time quantum algorithm for an \np-complete problem, is yet to 
be discovered! 

See \cite{aharonov:98} or www.qubit.org to learn more.

\subsection{Analytical solubility of Ising models}
\label{sec:ising}

Some problems in statistical physics have been exactly solved, but the
majority of problems has withstand the efforts of generations of
mathematicians and physicists. Why are some problems analytically
solvable whereas others, often similar, are not? Relating this
question to the algorithmic complexity of evaluating the partition
function gives us no final answer but helps to clarify the borderline
that separates the analytically tractable from the intractable
problems.

\begin{figure}[htb]
  \begin{center}
    \includegraphics[width=\columnwidth]{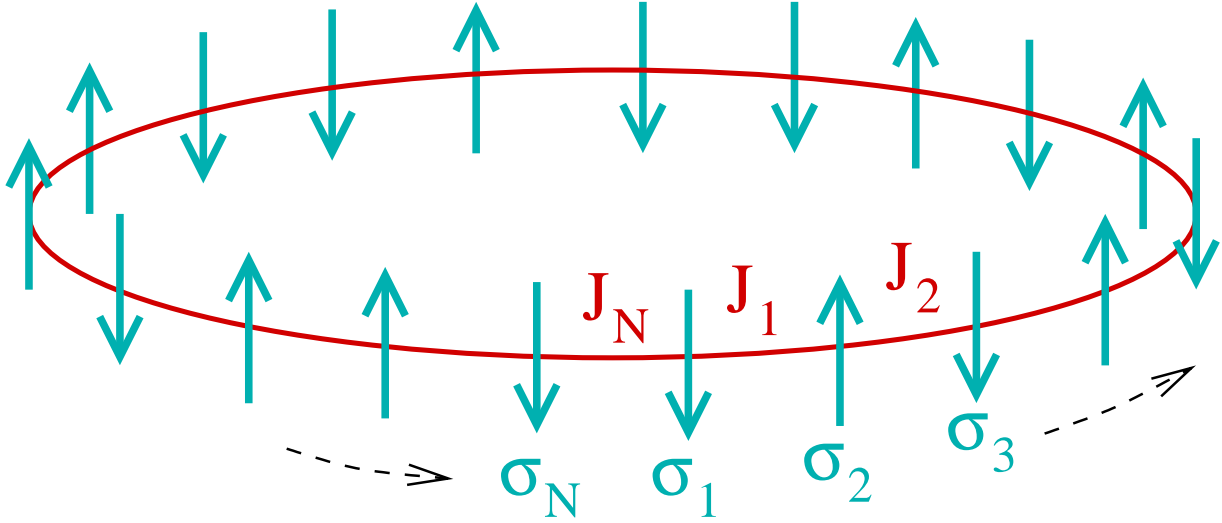}
  \end{center}
  \caption[Fig1]{\label{fig:spinring}Onedimensional Ising spin glass with 
    periodic boundary conditions.
    The partition sum of this system can be calculated in polynomial
    time.}
\end{figure}

As an example consider the Ising spin glass on a general graph $G$ \cite{hayes:spin}.
Let $\sigma=(\sigma_1,\ldots,\sigma_N)$ be an assignment of Ising
spins $\sigma_i=\pm1$ (``up'' or ``down''). 
The energy of a configuration $\sigma$ is
\begin{equation}
  \label{eq:ising-on-graph}
  E(\sigma) = -\sum_{\langle i,j\rangle}J_{i,j}\sigma_i\sigma_j - H\sum_i\sigma_i,
\end{equation}
where $H$ is the external magnetic field, $J_{i,j}$ are the coupling
constants and the first summation is over all edges in $G$. The
fundamental problem in statistical mechanics is to determine the
partition function
\begin{equation}
  \label{eq:Z-ising-on-graph}
  Z_N(G) = \sum_\sigma e^{-\beta E(\sigma)}.
\end{equation}
Evaluating the sum directly requires $\bigo(2^N)$ operations.  The
notion ``analytic solution'' has no precise definition, but as a
minimum requirement we want this number to be reduced from being
exponential to being polynomial in $N$. As an example consider the
well known transfer matrix solution of the one dimensional Ising glass
with periodic boundary conditions and coupling $J_k$ between spins
$\sigma_k$ and $\sigma_{k+1}$ (Fig.~\ref{fig:spinring}),
\begin{equation}
  \label{eq:Z-ising-1d}
  Z_N(\mathrm{ring}) = \mathrm{Tr}\prod_{k=1}^N\left(
  \begin{array}{cc}
      e^{\beta(J_k+H)} & e^{-\beta J_k} \\
      e^{-\beta J_k} & e^{\beta(J_k-H)}
  \end{array}
  \right),
\end{equation}
which can be evaluated in $\bigo(N)$ elementary operations. Since any
exact solution must include all numbers $J_k$, this is the best we can
expect. In the homogeneous case $J_k\equiv J$, where we can calculate
the product of transfer matrices,
\begin{equation}
  \label{eq:1d-ising-sol2}
  Z_N = \lambda_+^N + \lambda_-^N \mbox{ with }
  \lambda_{\pm} = e^{\beta J}\left[
  \cosh(\beta H) \pm \sqrt{\cosh^2(\beta H)-2 e^{-2\beta H}\sinh(2\beta H)}
  \right]
\end{equation}
the evaluation complexity drops to $\bigo(\log N)$ using fast
exponentiation.

Writing the partition sum as
\begin{equation}
  \label{eq:Z-counting}
  Z(G) = \sum_{E_k} n(E_k) e^{-\beta E_k}
\end{equation}
where the sum is over all possible energy values, it becomes obvious
that calculating $Z$ is closely related to the \#\p problem of
determining $n(E_k)$.  For general finite graphs $G$, this problem has
proven to be \#\p-complete
\cite{jerrum:sinclair:90,jaeger:vertigan:welsh:90}, so there is no
hope to find an analytical solution (even in the weak sense above).
This situation hardly improves if we restrict the graphs to the more
``physical'' crystal lattices: computing the partition function for a
finite sublattice of a {\em non-planar crystal lattice} is
\#\p-complete \cite{istrail:2000}.  This includes every crystal lattice
in $d > 2$, the d=2 model with next-nearest neighbor interactions, two
coupled $d=2$ models etc..  It also includes all models with $d\geq 2$
and external field $H$, since these can be transformed into zero-field
models on an augmented graph $\hat{G}$ which is non-planar unless the
underlying lattice graph $G$ is onedimensional. The construction of
$\hat{G}$ from $G$ is easy, just adjoin an additional vertex  
(spin) $\sigma_0$ to $G$ and let the additional edges $\sigma_0\sigma_i$
have constant interaction energy $J_{0,i}=H$.  The partition function
of the augmented system reads
\begin{equation}
  \label{eq:ising-augmented}
  Z(\hat{G}) = \sum_{\sigma}e^{-\beta[-\sum J_{i,j}\sigma_i\sigma_j]}\Big(
  e^{-\beta H\sum\sigma_i}+e^{\beta H\sum\sigma_i}\Big),
\end{equation}
where the additional factor comes from the new spin $\sigma_0=\pm1$.
From this expression it is easy to see that $Z(\hat{G})$ equals two
times the partition function $Z(G)$ in field $H$.

But where are the soluble Ising models? It has been proven that the
partition function of Ising systems on {\em planar crystal lattices}
can be evaluated in polynomial time \cite{barahona:82}. This includes
the celebrated Onsager solution of the square ferromagnet
\cite{onsager:44} as well as the onedimensional example from above.
It turns out that the Ising model's partition sum can be calculated in
polynomial time for {\em all graphs of fixed genus $g$}
\cite{regge:zecchina:00, galluccio:loebl:vondrak:00}. A graph has
genus $g$ if it can be drawn on an orientable surface of genus g (a
sphere with $g$ ``handles'' attached to it) without crossing the
edges.  Planar graphs have genus $0$, toroidal graphs have genus $1$
and so on.  For the crystal lattices in $d > 2$, the genus increases
monotonically with the number of spins, i.e.\ it is not fixed
\cite{regge:zecchina:00}.

The mechanism for proving tractability or intractability is the same
in statistical mechanics as in computational complexity: polynomial
reduction. Barahona \cite{barahona:82} for example applies a reduction
of the \np-hard problem \problem{Max Cut} to the Ising spin glass in $3d$
to proof the \np-hardness of the latter. A reduction of the planar
Ising model to \problem{Minimum Weight Matching} on the other hand proofs
the tractability of the former, since \problem{Minimum Weight Matching}
can be solved in polynomial time.

Note that in our classification of spin systems 
the nature of the
couplings is not significant. A frustrated, glassy system with random
couplings $J_{i,j}$ of both signs is in the same class as the
homogeneous ferromagnet with $J_{i,j} \equiv J > 0$ as long as the
underlying graph $G$ is the same.  In our one\-di\-men\-sion\-al example we
did not discriminate these cases: they are both polynomial. This
situation changes of course if we consider the ground states rather
than the complete partition function. Here the nature of the couplings
matters a lot: finding the ground states in pure ferromagnetic systems
is trivial on all lattices, whereas it is \np-hard for glassy systems
with positive and negative couplings on non-planar lattices
\cite{barahona:82}.

A lot of other problems arising in statistical physics can be
classified according to the computational complexity to evaluate their
partition function, see \cite{welsh:90} and references therein. It
turns out that all the problems known to have an exact solution
\cite{baxter:82} can be evaluated in polynomial time.  Problems that
are \#\p-complete, however, are very unlikely to be exactly solvable.
Anyone looking for an exact solution of such a problem should keep in
mind, that he or she is simultaneously attacking \problem{TSP}, 
\problem{Hamiltonian cycle}, \problem{SAT} and all the other NP-hard
problems. In statistical mechanics the focus is on results for the
thermodynamic limit $N\to\infty$ rather than for finite systems,
however.  It is not clear how much of the ``hardness'' survives in
this limit.

\subsection{Probabilistic analysis of combinatorial problems}
\label{sec:prob-analysis}

Problems from combinatorial optimization can formally considered as
models in statistical mechanics. The cost function is renamed
Hamiltonian, random instances are samples of quenched disorder, and
the ground states of the formal model correspond to the solutions of
the optimization problems.  In this way methods developed in the
framework of statistical mechanics of disordered systems become
powerful tools for the probabilistic analysis of combinatorial
problems \cite{mezard:etal:87}.

Statistical mechanics methods have been applied for example to the
\problem{TSP} \cite{mezard:parisi:86,krauth:mezard:89},
\problem{Graph Partitioning} \cite{fu:anderson:86} and \problem{$k$-SAT}
\cite{kirkpatrick:selman:94,monasson:zecchina:97}.  A particular nice
example of this approach is the analysis of \problem{Assignment} (also
called \problem{Bipartite Matching}): Given an $N\times N$ matrix with
non-negative entries $a_{i,j}\geq 0$.  Find
\begin{equation}
  \label{eq:ap-costs}
  E^*_N = \min_{\sigma} \sum_{i=1}^N a_{i,\sigma(i)}
\end{equation}
where the minimum is taken over all permutations $\sigma$ of
$(1,\ldots,N)$. 

A probabilistic analysis aims at calculating average properties for an
ensemble of random instances, the canonical ensemble being random
numbers $a_{i,j}$ drawn independently from a common probability
density $\rho(a)$. Using the replica method from statistical physics,
Marc M\'ezard and Giorgio Parisi \cite{mezard:parisi:85} found (among other things)
\begin{equation}
  \label{eq:ap-replica}
  \lim_{N\to\infty}\disave{E^*_N} = \frac{\pi^2}{6},
\end{equation}
where $\disave{\cdot}$ denotes averaging over the $a_{i,j}$.  
A rigorous proof of eq.~\ref{eq:ap-replica} has been presented recently
\cite{aldous:00} and represents one of the rare cases where a
replica result has been confirmed by rigorous methods.

Some years after the replica-solution, Parisi recognized that for
exponentially distributed matrix elements ( $\rho(a)=e^{-a}$) the
average optimum for $N=1$ and $N=2$ is
\begin{equation}
  \label{eq:n12-exponential}
  \disave{E^*_1} = 1 \qquad \disave{E^*_2} = 1+\frac{1}{2^2}.
\end{equation}
From this and the fact that the replica result for $N\to\infty$ can be
written as
\begin{equation}
  \label{eq:pi-series}
  \frac{\pi^2}{6} = \sum_{k=1}^\infty \frac{1}{k^2}
\end{equation}
he conjectured \cite{parisi:98} that the average optimum for finite
systems is
\begin{equation}
  \label{eq:parisis-conjecture}
   \disave{E^*_N} = \sum_{k=1}^N \frac{1}{k^2}.
\end{equation}
Parisi's conjecture is supported by numerical simulations, but no
formal proof has been found despite some efforts
\cite{dotsenko:00,coppersmith:sorkin:99}.

Note that Eqs.~\ref{eq:n12-exponential} and
\ref{eq:parisis-conjecture} only hold for $\rho(a)=e^{-a}$, whereas
eq.~\ref{eq:ap-replica} is valid for all densities with $\rho(a\to 0)
= 1$. For the uniform density on $[0,1]$ the first terms are
\begin{equation}
  \label{eq:n12-uniform}
  \disave{E^*_1} = \frac{1}{2} \qquad \disave{E^*_2} = \frac{23}{30}.
\end{equation}
If you can you guess the expression for general, finite $N$ in this case,
please send me an email.

Sometimes a statistical mechanics analysis not only yields
exact analytical results but also reveals features that are
important to design and understand algorithms. 
A recent example is the analysis of the Davis-Putnam-Algorithm
for \problem{SAT} \cite{cocco:monasson:01a,cocco:monasson:01b}.
Another example is given
by the number partitioning problem \problem{NPP} \cite{hayes:npp}, an \np-hard
optimization problem, where it has been shown, that for this particular 
problem no heuristic approach can be better than stupid random
search \cite{mertens:00a,mertens:01,borgs:chayes:pittel:01}.

\subsection{Phase transitions in computational complexity}
\label{sec:phase-transition}

The theory of computational complexity is based entirely on worst-case
analysis. It may happen that an algorithm requires exponential time on
pathological instances only.  A famous example is the Simplex Method
for \problem{Linear Programming}.  Despite its exponential worst-case
complexity it is used in many applications to solve really large
problems. Apparently the instances that trigger exponential running
time do not appear under practical conditions.

Now \problem{Linear Programming} is in \p\ thanks to the Ellipsoid algorithm
\cite{papadimitriou:steiglitz:82}, but similar scenarios are observed
for \np-hard problems. Take \problem{$3$-SAT} as an example.  Generating
random instances with $N$ variables and $M$ clauses and feeding them
to a smart but exhaustive algorithm one observes polynomial running
time {\em unless} the ratio $M/N$ is carefully adjusted. If $M/N$ is
too low, the problem is {\em underconstrained}, i.e.\ has many
satisfying solutions, and a clever algorithm will find one of these
quickly. If $M/N$ is too large, the problem is {\em overconstrained},
i.e.\ has a large number of contradictory constraints which again a
clever algorithm will discover quickly \cite{hayes:97}. The real hard
instances are generated for ratios $\alpha = M/N$ close to a critical
value $\alpha_c$ \cite{monasson:etal:99}.
  
The transition from underconstrained to overconstrained formulas in in
\problem{$3$-SAT} bears the hallmarks of a phase transition in physical
systems.  The control parameter is $\alpha$, the order parameter is
the probability of the formula being satisfiable. Similar phase
transitions have been discovered and analyzed with methods from
statistical mechanics in other computational problems. See the special
issue of {\em Theoretical Computer Science} on ``Phase Transitions in
Combinatorial Problems'' for a recent overview of this interdisciplinary
field \cite{tcs-phasetransitions}.

%%% Local Variables: 
%%% mode: latex
%%% TeX-master: "lecture"
%%% End: 

\bibliographystyle{unsrt}

\bibliography{complexity,cs,math,quantum}

\end{document}